\documentclass{emulateapj}
\usepackage{amsmath}
\usepackage{amssymb}
\usepackage{natbib}
\usepackage{mwe}
\usepackage{graphicx}
\usepackage{color}
\usepackage{hyperref}

\shorttitle{Declining rotation curves in $\Lambda$CDM}
\shortauthors{A. F. Teklu et al.}

\begin{document}

\title{Declining rotation curves at $z=2$ in $\Lambda$CDM galaxy formation simulations}

\author{Adelheid F. Teklu$^{1,2}$, Rhea-Silvia Remus$^{1,3}$, Klaus Dolag$^{1,4}$, Alexander Arth$^{1,5}$, \\Andreas Burkert$^{1,5}$, Aura Obreja$^1$, \& Felix Schulze$^{1,5}$}
\affil{$^1$ Universit\"ats-Sternwarte M\"unchen, Scheinerstra{\ss}e 1, D-81679 M\"unchen, Germany\\
$^2$ Excellence Cluster Universe, Boltzmannstra{\ss}e 2, D-85748 Garching, Germany\\
$^3$ Canadian Institute for Theoretical Astrophysics, 60 St. George Street, University of Toronto, Toronto ON M5S 3H8, Canada \\
$^4$ Max-Planck Institute for Astrophysics, Karl-Schwarzschild-Str. 1, D-85741 Garching, Germany\\
$^5$ Max-Planck Institute for Extraterrestrial Physics, Giessenbachstra{\ss}e 1, D-85748 Garching, Germany \\ \texttt{ateklu@usm.lmu.de} \\}


\begin{abstract}

Selecting disk galaxies from the cosmological, hydrodynamical simulation {\it Magneticum Pathfinder} we show that
almost half of our poster child disk galaxies at $z=2$ show significantly declining rotation curves and low
dark matter fractions, very similar to recently reported observations. 
These galaxies do not show any anomalous behavior, reside in standard dark matter halos and typically grow significantly in 
mass until $z = 0$, where they span all morphological classes, 
including disk galaxies matching present day rotation curves and observed dark matter fractions. 
Our findings demonstrate that declining rotation curves and low dark
matter fractions in rotation dominated galaxies at $z=2$ appear naturally within the $\Lambda$CDM paradigm and
reflect the complex baryonic physics, which plays a role at the peak epoch of star-formation. 
In addition, we find some dispersion dominated galaxies at $z=2$ which host a significant gas disk
and exhibit similar shaped rotation curves as the disk galaxy population, 
rendering it difficult to differentiate between these two populations with currently available observation techniques. 

\end{abstract}



\section{Introduction}

Since the postulation of dark matter (DM) by \citet{Zwicky33}, many observational studies analyzing rotation curves of galaxies \citep[e.g.][]{Rubin78} have supported this picture:
While rotational velocities ($V^\mathrm{rot}$) deduced from the visible matter should decrease proportional to  $r^{-1/2}$ in the outer parts of galaxies, they were found to remain flat. 
The knowledge of this discrepancy in the mass content and thus the need for an explanation for this missing mass lead to the acceptance of dark matter as the dominant mass component of galaxies (see \citet{Naab17} for a detailed review).

Recently, \citet{Genzel17} \citep[see also][]{Lang17} presented measurements of rotation curves at redshift $z\approx2$ that do not stay flat but decrease with increasing radius, opening a debate about the importance and even presence of DM in the outer disks and inner halos of these massive systems (and generally at higher redshift). 
In this letter we investigate whether the existence of decreasing rotation curves at high redshifts contradicts or actually is a natural outcome of the $\Lambda$CDM paradigm, using the state-of-the-art cosmological simulation Magneticum Pathfinder\footnote{www.magneticum.org} (K. Dolag et al., in preparation).


\begin{figure*}
    \includegraphics[width=0.24\textwidth]{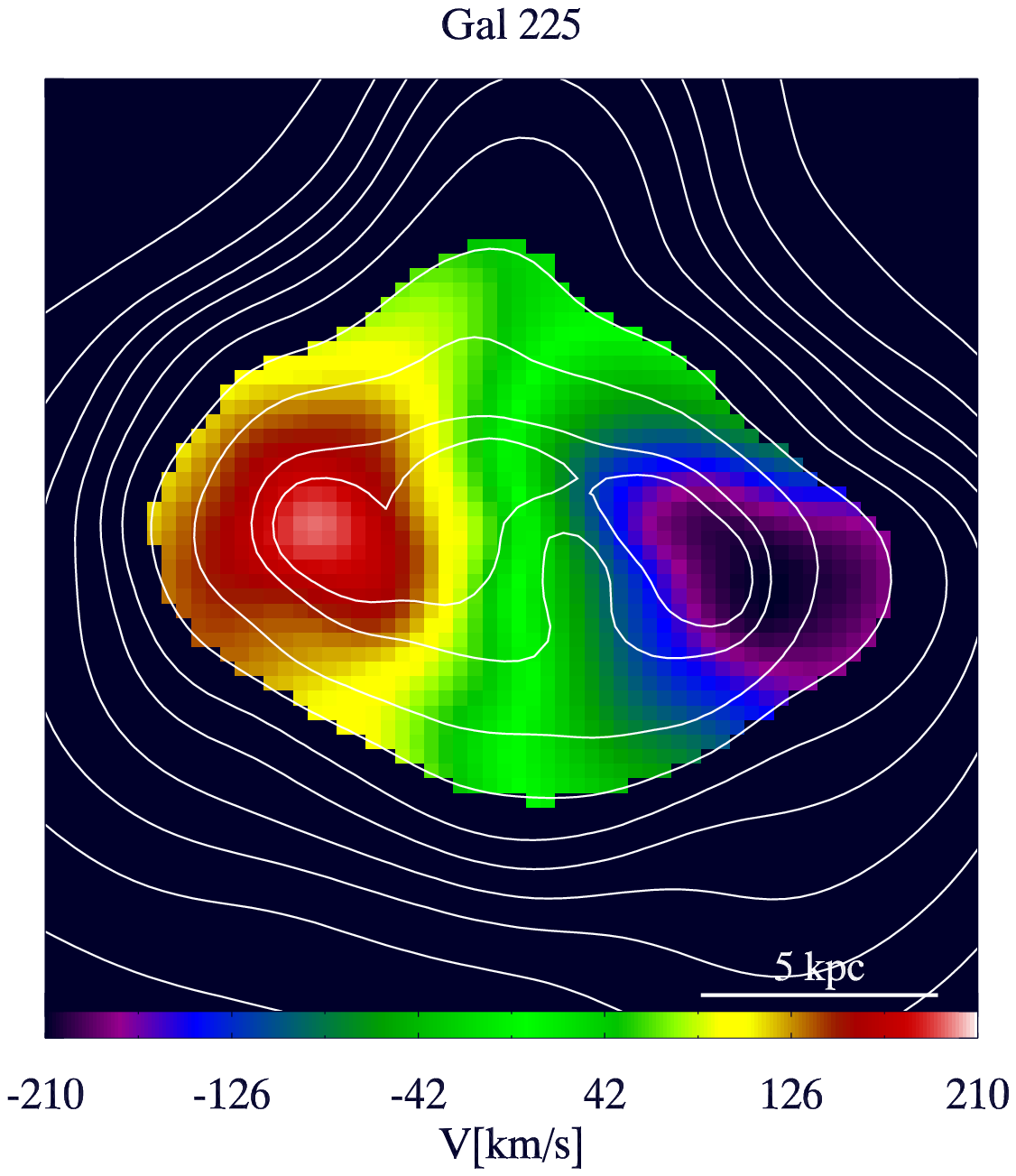}
    \includegraphics[width=0.24\textwidth]{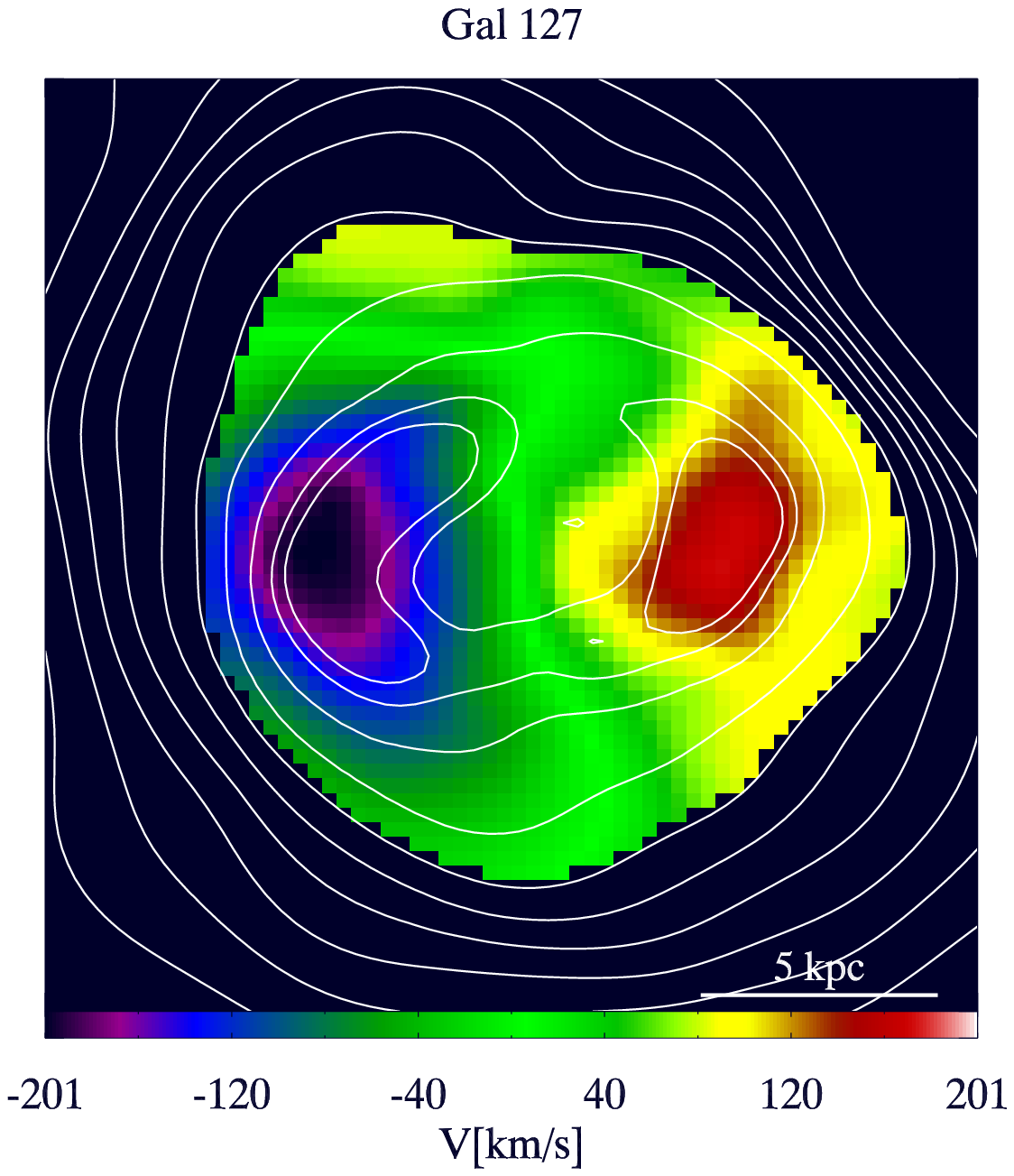}
    \includegraphics[width=0.24\textwidth]{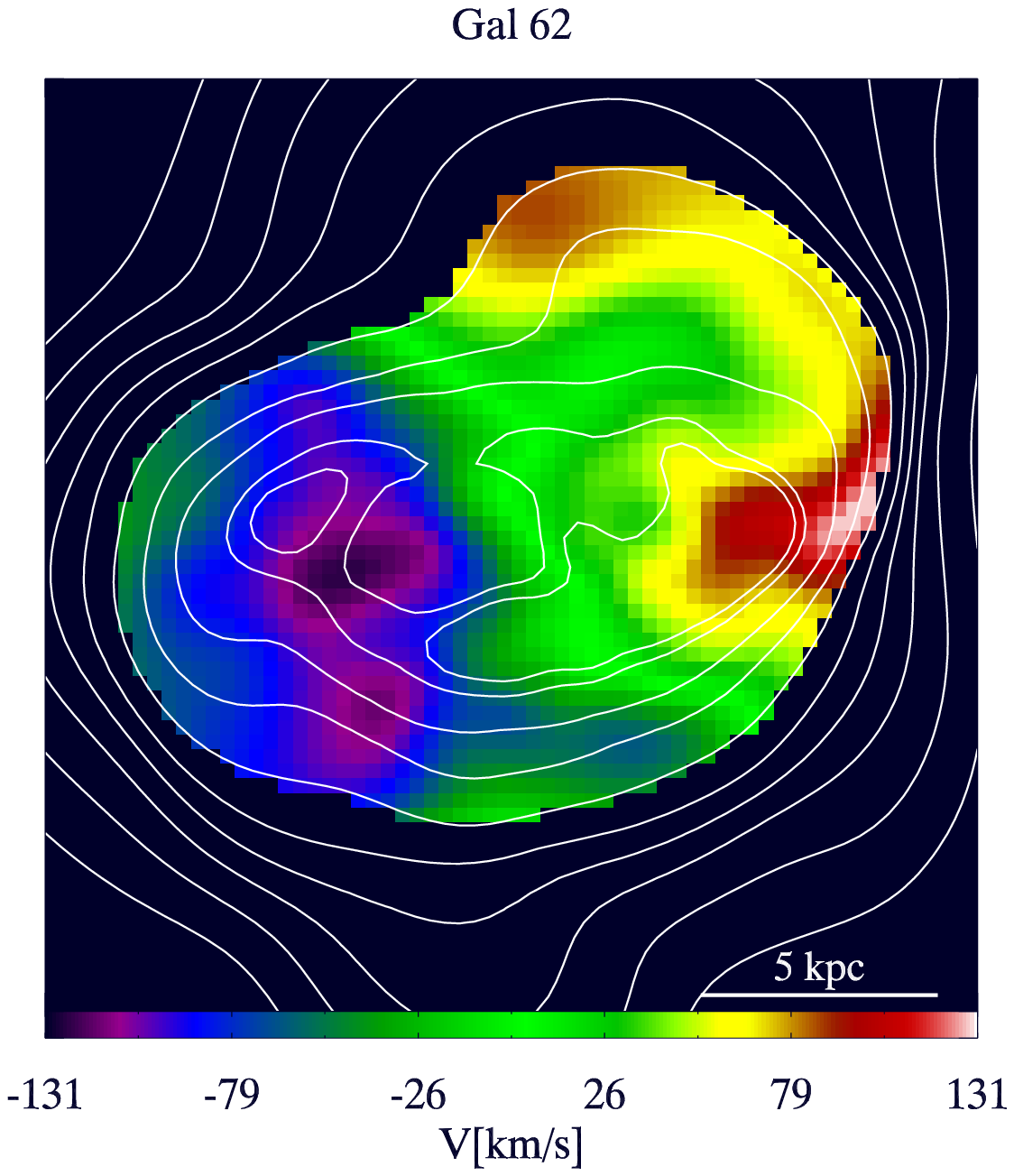}
    \includegraphics[width=0.24\textwidth]{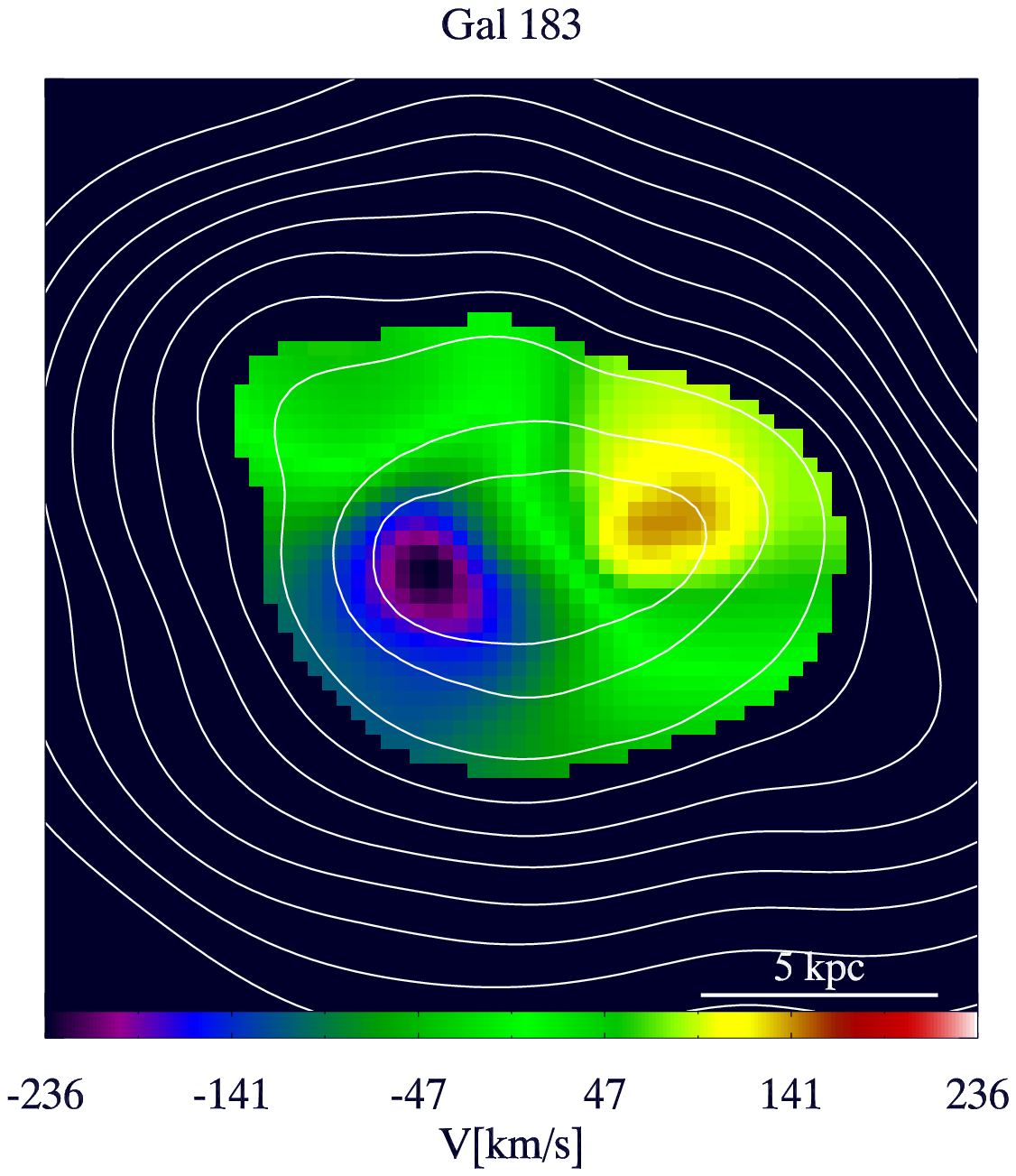}\\
    
    \includegraphics[width=0.24\textwidth]{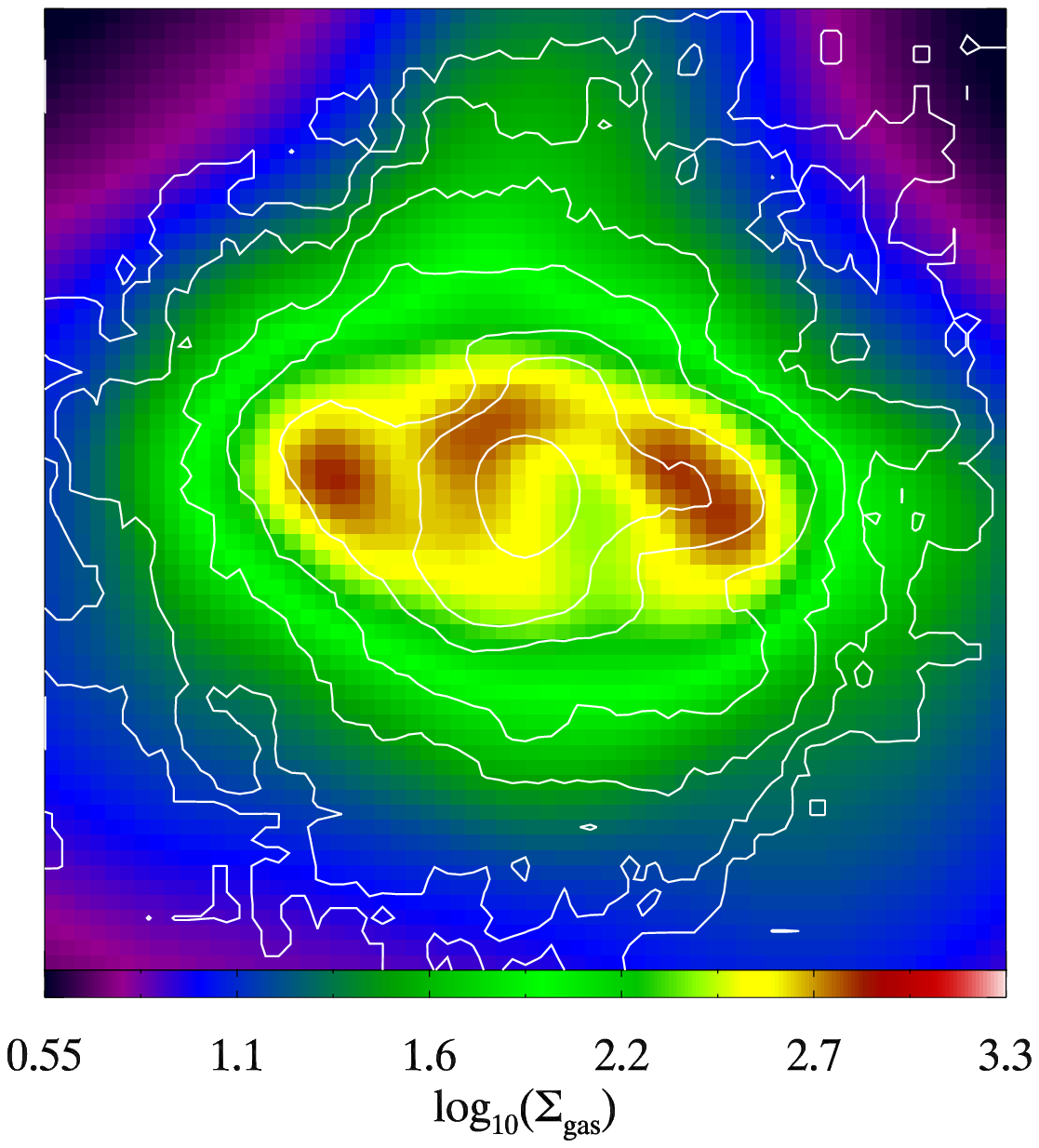}
    \includegraphics[width=0.24\textwidth]{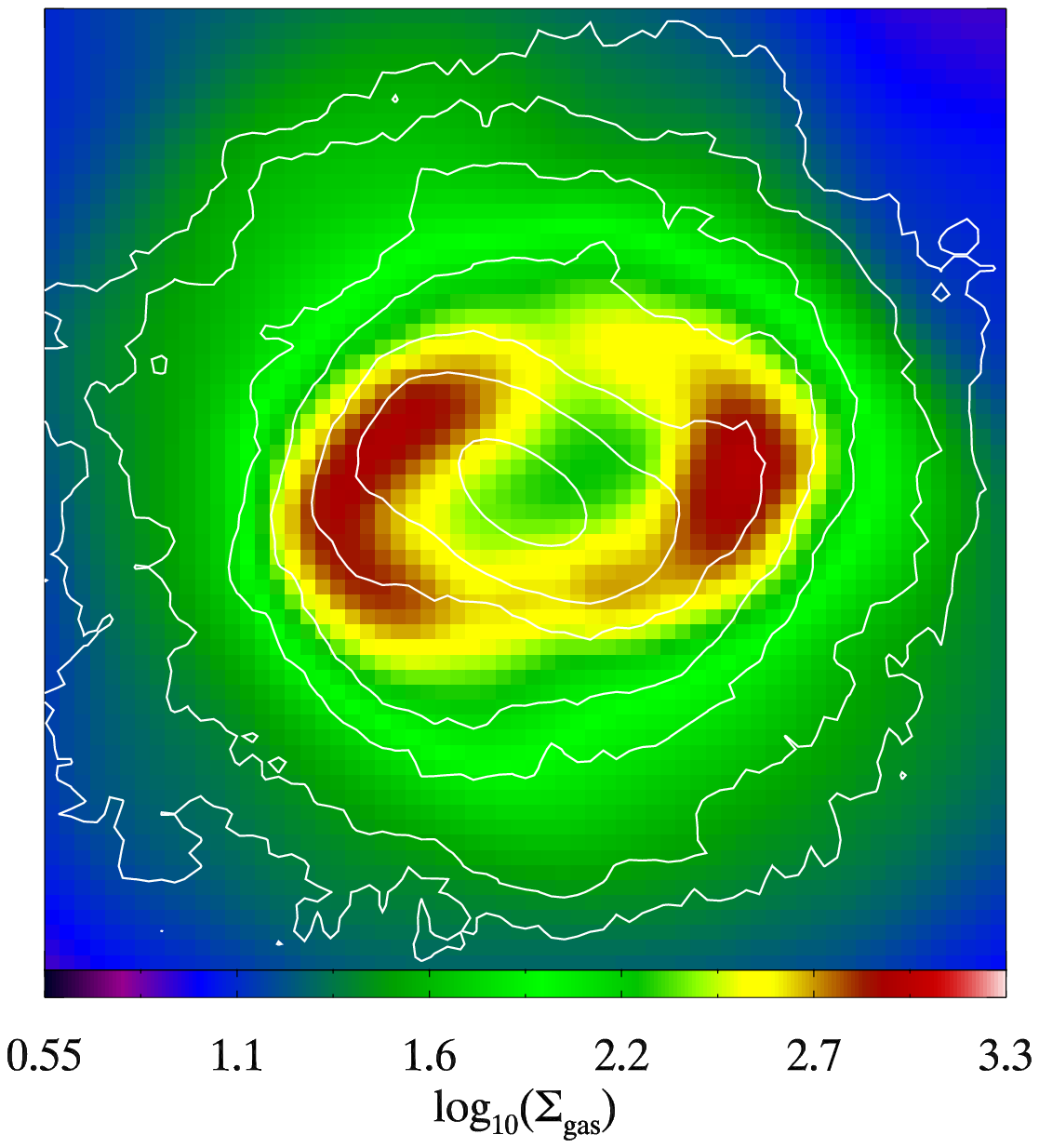}
    \includegraphics[width=0.24\textwidth]{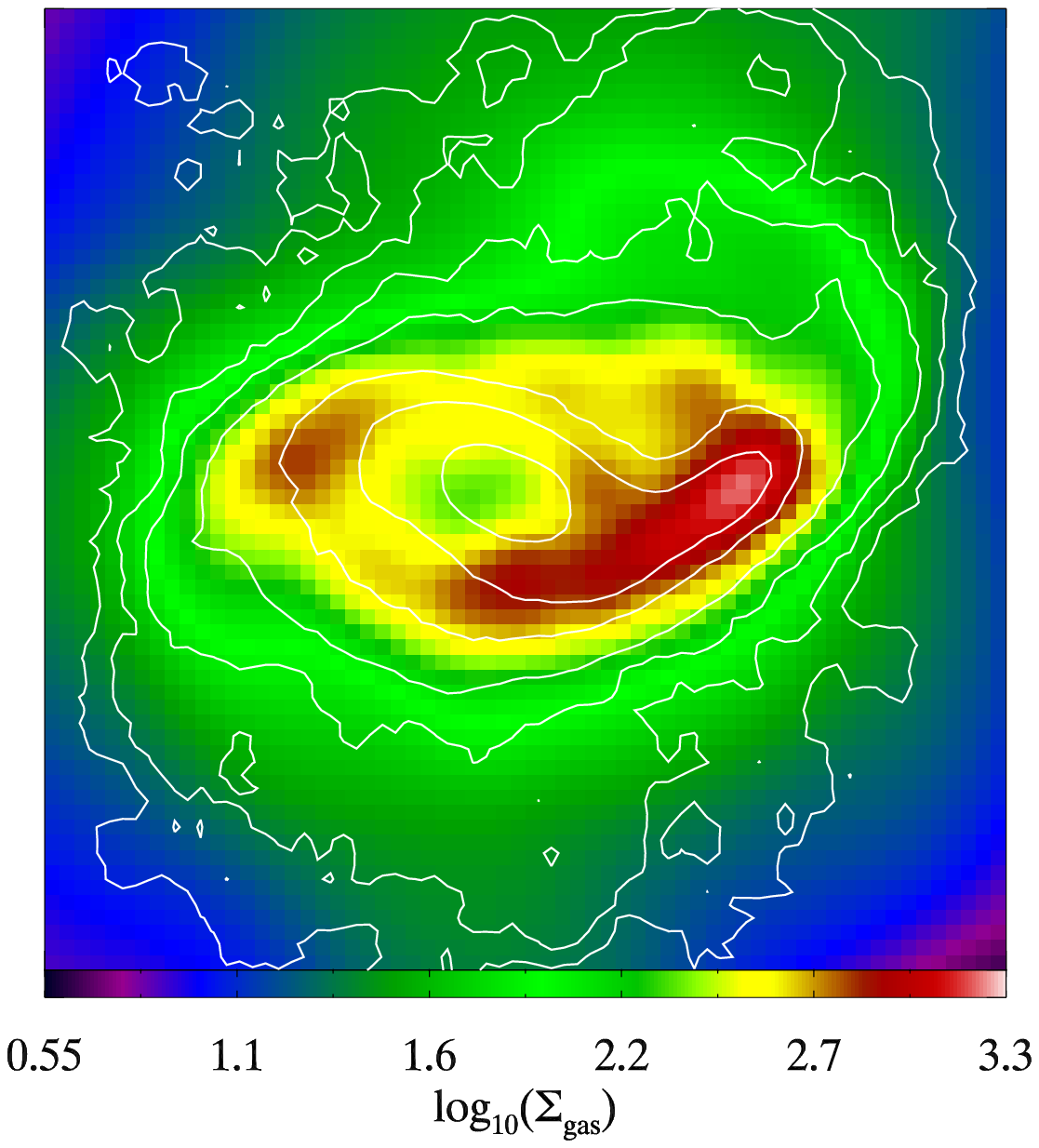}
    \includegraphics[width=0.24\textwidth]{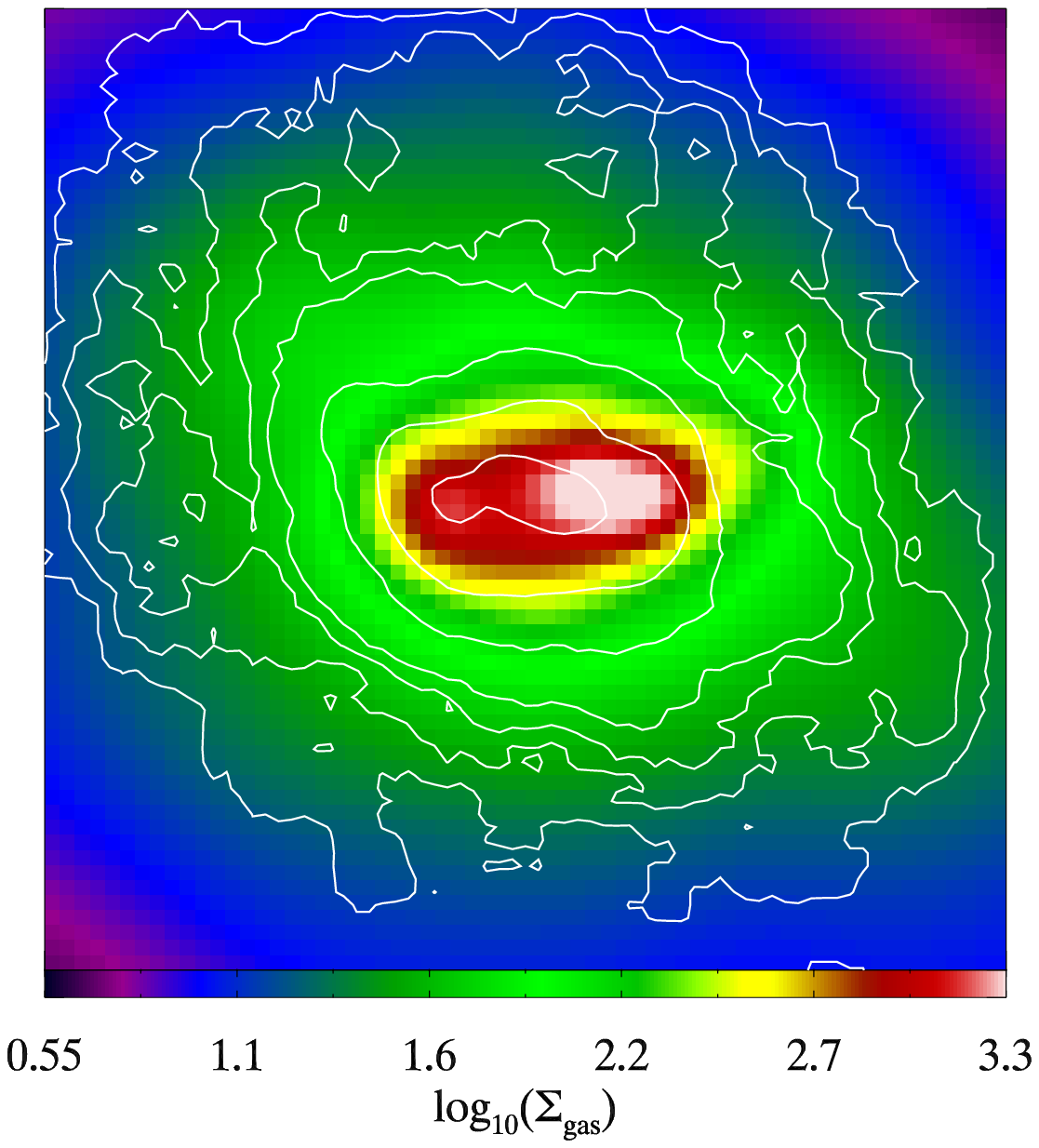} \\
    
    \includegraphics[width=0.24\textwidth,clip=true]{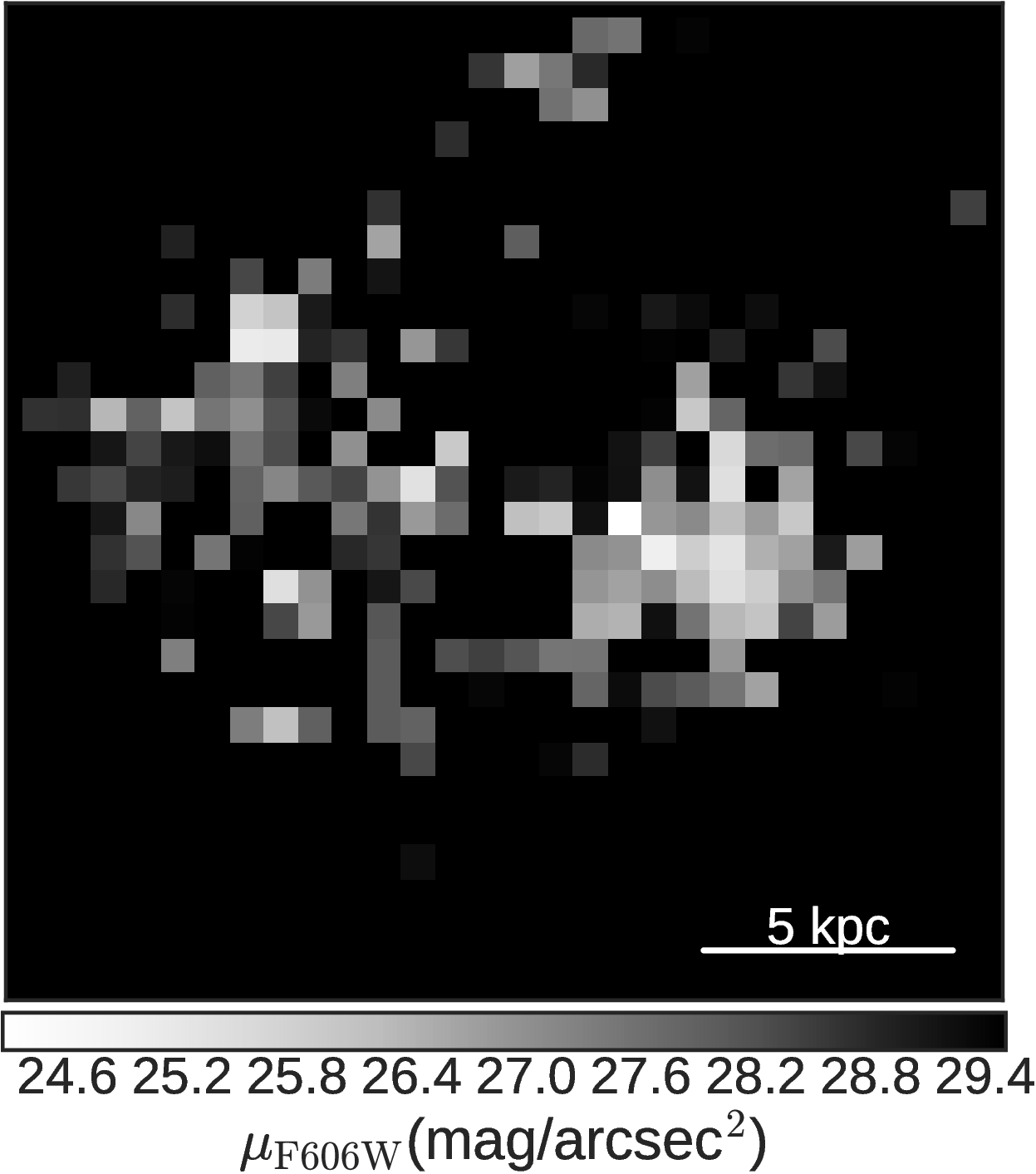}
    \includegraphics[width=0.24\textwidth,clip=true]{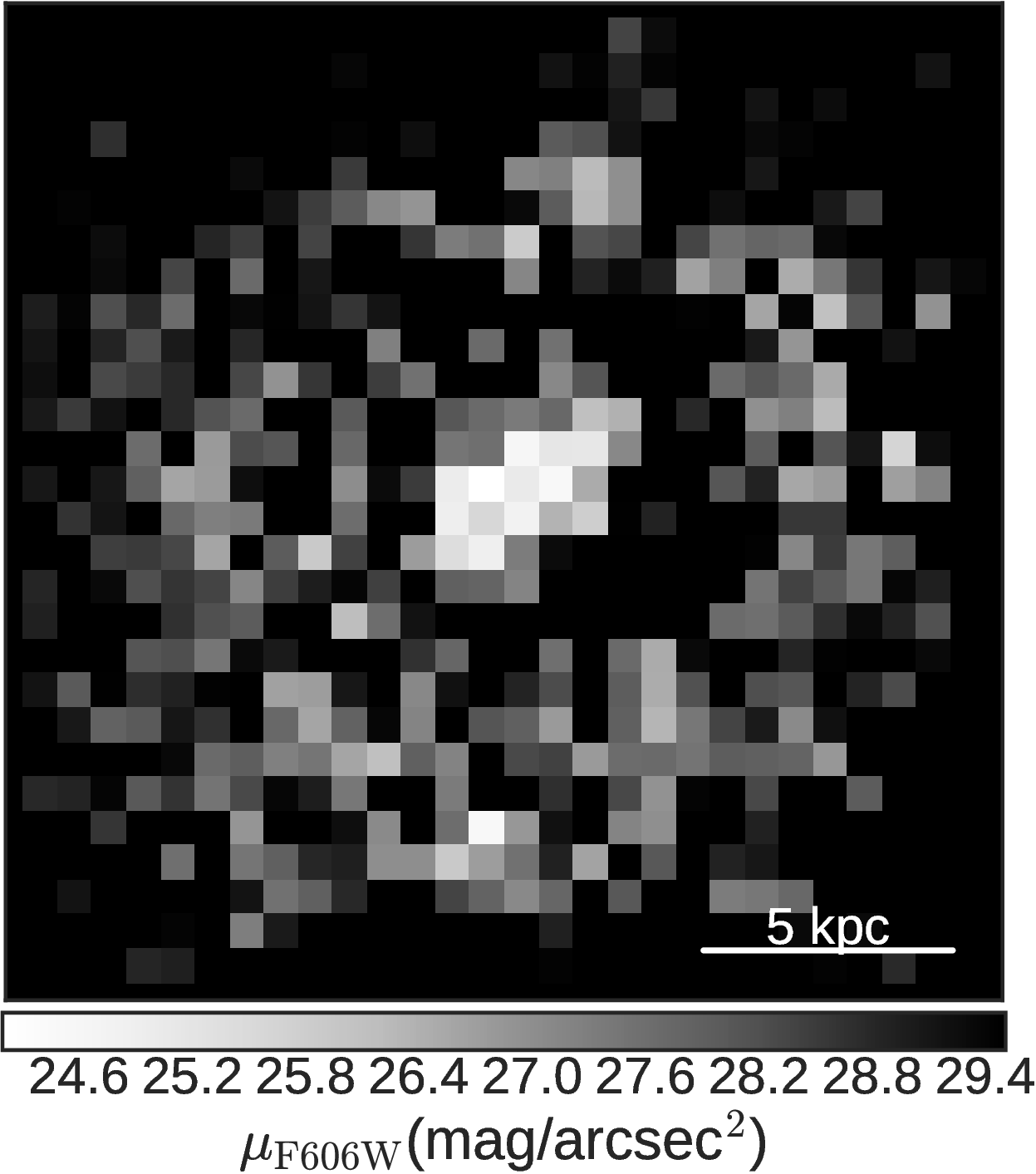}
    \includegraphics[width=0.24\textwidth,clip=true]{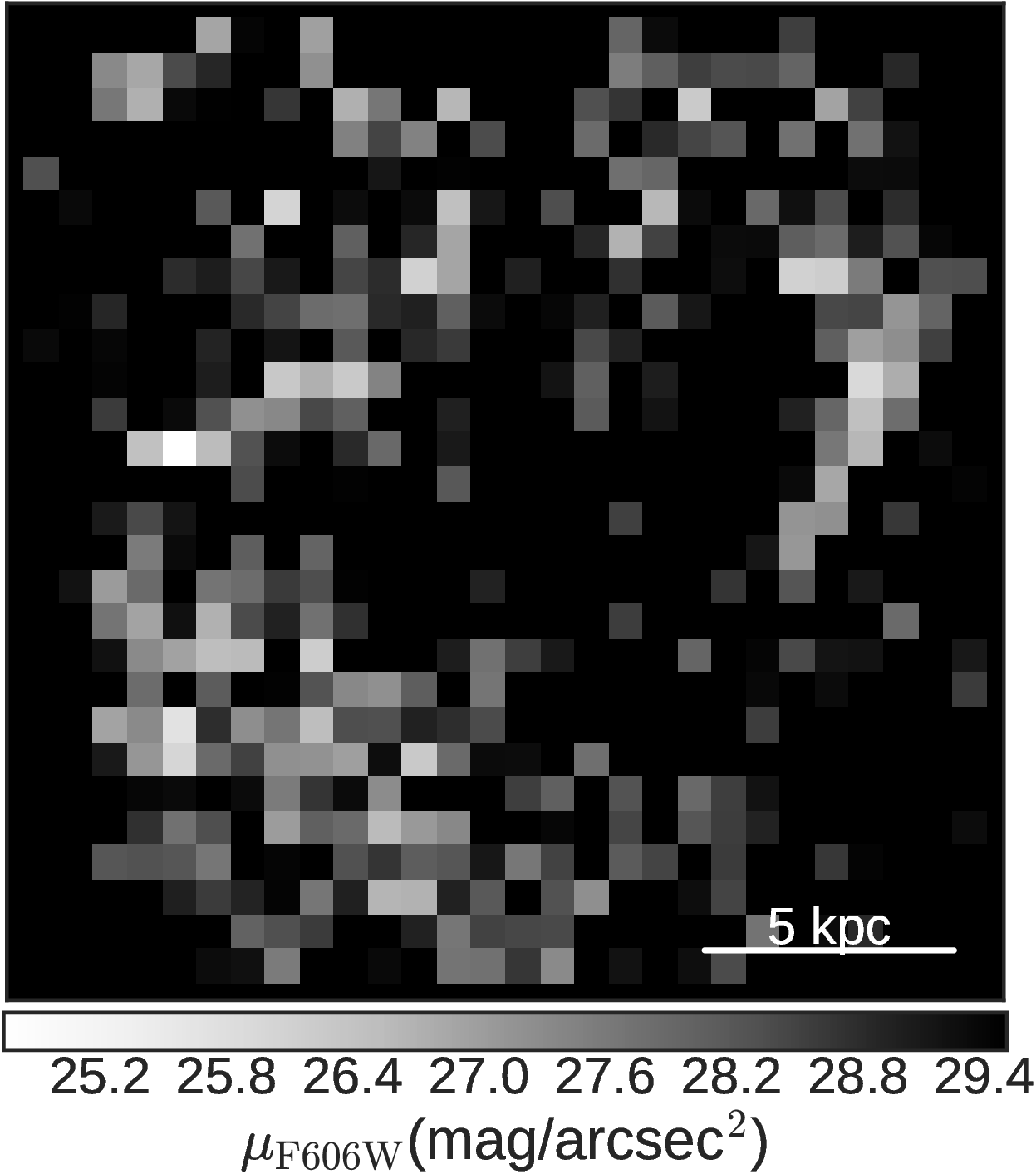}
    \includegraphics[width=0.245\textwidth,clip=true]{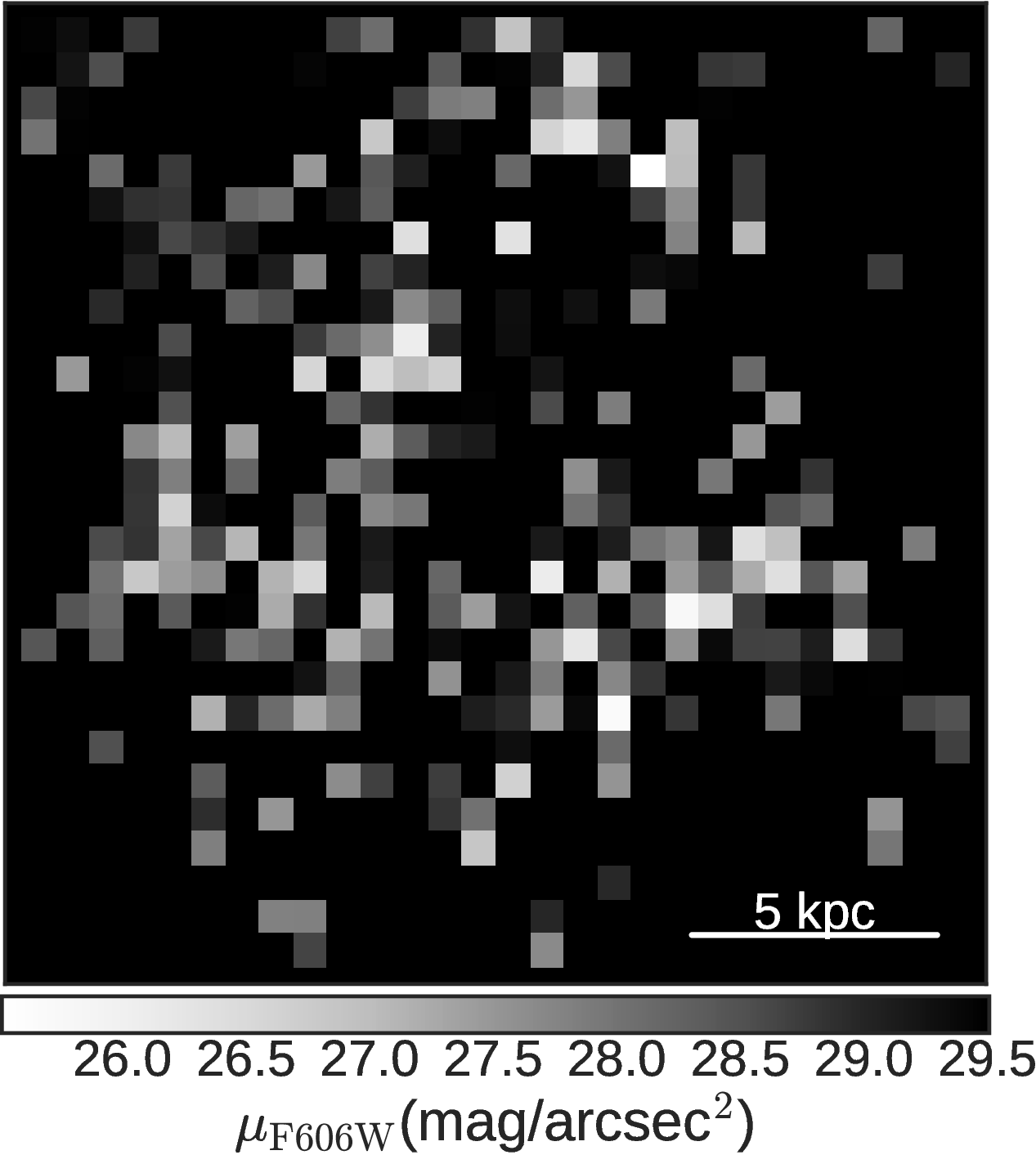}
    
    \caption{Example galaxies from the $z=2$ sample with declining rotation curves (see Fig. \ref{fig:rotcurve2}), from left to right the three disk galaxies {\it gal~225},
    		{\it gal~127}, and {\it gal~62}, and the gas-rich spheroidal system 
    		{\it gal~183}, rotated to inclinations (e.g. $i=60$, $i=45$, $i=25$
      and $i=75$, respectively) similar to those of the galaxies presented in \citet{Genzel17}.
            \textit{Upper row:} Velocity maps of the cold gas component for each galaxy, with contours of the cold gas column density overlayed.
            \textit{Middle row:} Cold gas column density maps with overlayed stellar column density contours.
            \textit{Lower row:} Simulated  HST broadband F606W images using {\em GRASIL-3D}.}
    \label{fig:maps}
\end{figure*}

\section{The Simulations}\label{sec:sim}
The {\it Magneticum Pathfinder} simulations are a set of state-of-the-art,
cosmological, hydrodynamical simulations \citep[see][for details on the numerical scheme]{Beck15} of
different cosmological volumes with different resolutions. They
follow a standard $\Lambda$CDM cosmology with parameters
($h$, $\Omega_{M}$, $\Omega_{\Lambda}$, $\Omega_{b}$, $\sigma_{8}$) set to
($0.704$, $0.272$, $0.728$, $0.0451$, $0.809$), adopting a WMAP 7 cosmology \citep{Komatsu11}.

These simulations follow a wide range of physical processes \citep[see][for details]{Hirschmann14a,Teklu15}
which are important for studying the formation of active galactic nuclei (AGN), galaxies, and galaxy cluster.
The simulation set covers a huge dynamical range with a detailed treatment of key physical processes that are known to control galaxy evolution,
thereby allowing to reproduce the properties of the large-scale, intra-galactic, and intra-cluster medium \citep[see
e.g.][]{Dolag16,Gupta17,Remus17b} as well as the distribution of different chemical species within galaxies and galaxy clusters \citep{Dolag17}, and
the properties of the AGN population \citep{Hirschmann14a,Steinborn16}.
Especially, detailed properties of galaxies of different morphologies can be recovered and studied, for example their angular momentum properties and the evolution of the stellar mass--angular momentum relation with redshift \citep{Teklu15,Teklu16}, the mass-size relations and their
evolution \citep[see e.g.][]{Remus16,Remus17}, global properties like the fundamental plane \citep{Remus16} or dark matter fractions \citep{Remus17}, the baryon conversion efficiency \citep[see e.g.][]{Steinborn15,Teklu17}, as well as the dynamical properties of early type galaxies \citep{Schulze2018}.

For this study we use the simulation {\it Box4/uhr}, which covers a
volume of (68 Mpc)$^{3}$, initially sampled with $2\cdot576^{3}$
particles (dark matter and gas), leading to a mass resolution of
$m_\mathrm{gas} = 7.3\cdot10^{6}M_\odot$ for the gas and $m_\mathrm{stars} = 1.8\cdot10^{6}M_\odot$
for stellar particles, with a plummer equivalent gravitational softening corresponding
to $0.33$kpc at $z=2$ for the star particles and twice this value for the gas particles.


\begin{figure*}
 \begin{centering}
  \centerline{\includegraphics[width=0.49\textwidth,clip=true]{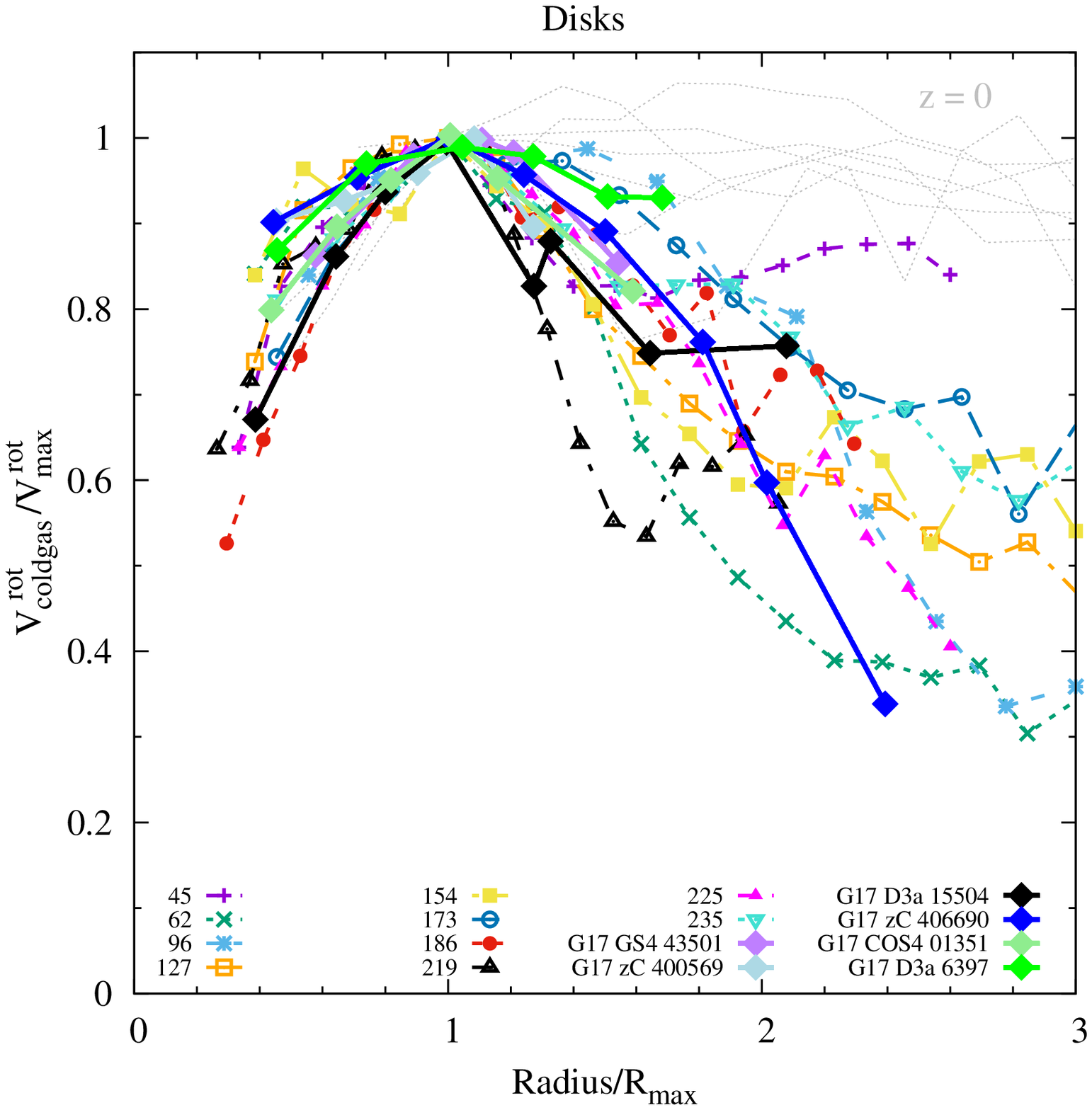} 
              \includegraphics[width=0.49\textwidth,clip=true]{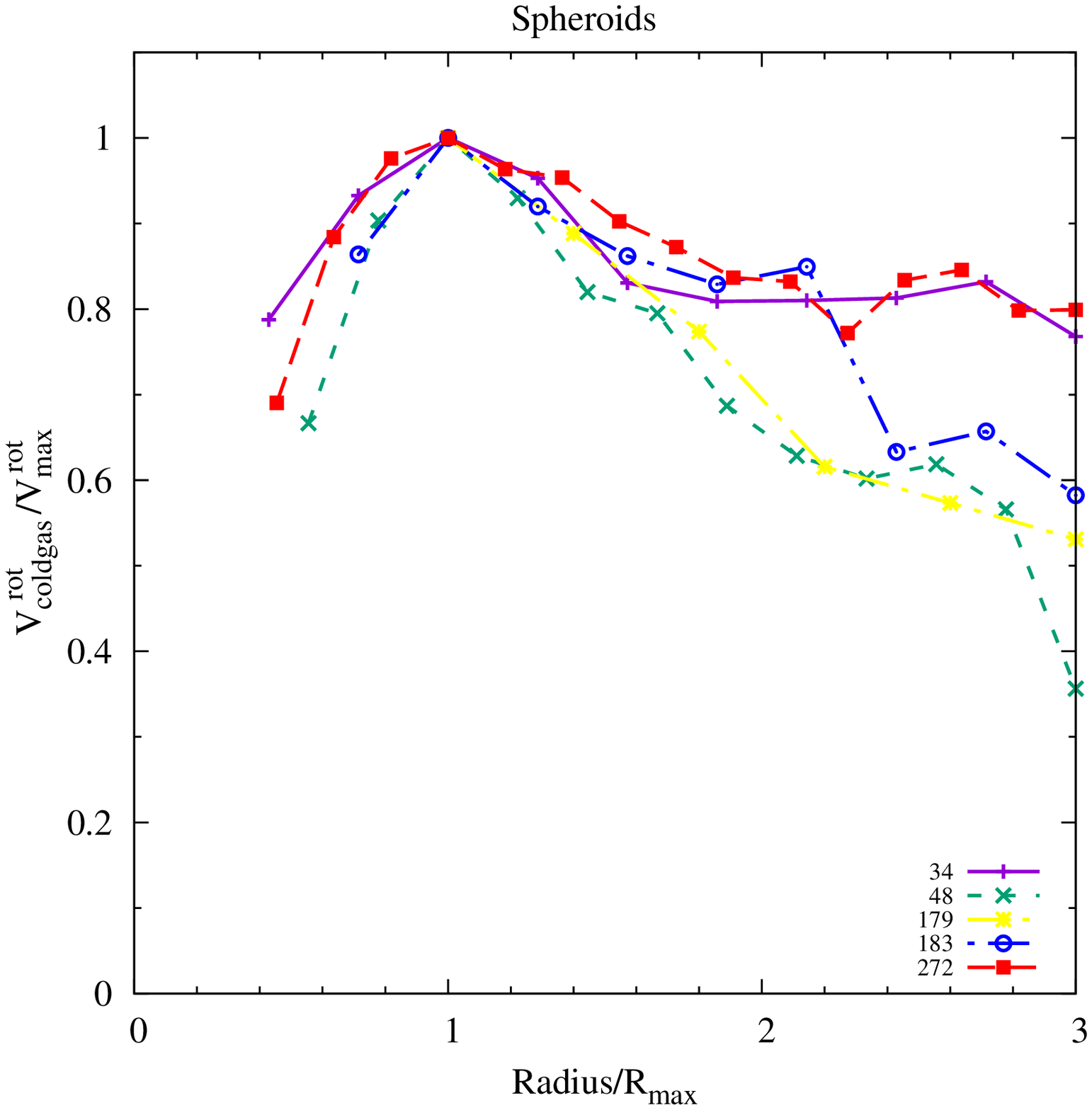}
            }
  \caption{Rotation curves obtained from the cold gas for 10 out of the 26 poster child disk galaxies which show clearly declining rotation curves (left panel) and for the 5 gas-rich spheroidal galaxies (right panel) at $z=2$, normalized by $V_\mathrm{coldgas}^\mathrm{rot}$ at the radius of maximum velocity $R_\mathrm{max}$. 
  The thick colored lines in the left panel show the 6 declining rotation curves presented in \citet{Genzel17}, while the gray lines show 7 poster child disk galaxies at $z=0$, using $\approx 1.4 \cdot R_{1/2}$ as $R_\mathrm{max}$.}
  \label{fig:rotcurve2}
 \end{centering}
\end{figure*}

\section{Sample of Galaxies}\label{sec:sample}
To ensure proper resolution of the inner structure, we only select halos with
virial masses above $5\cdot10^{11}M_{\odot}$ hosting galaxies with stellar masses
above $5\cdot 10^{10}M_{\odot}$ for this study. 
These mass ranges are consistent with the observed properties of the high-z galaxy sample of \citet{Genzel17}.
This leads to a sample of 212 and 275 halos at $z=2$ and $z=0$, respectively. 
Furthermore, we classifiy the galaxies based on the distribution of the
circularity parameter $\varepsilon=j_z/r\sqrt{GM(r)/r}$ of the stars
within the galaxies, where $j_{z}$ is the $z$-component
of the stars' specific angular momentum \citep[see also][]{Abadi03,Scan08}.
Thus, dispersion-dominated systems represent observed early-type galaxies and 
are characterized by a broad peak in the
distribution at $\varepsilon \simeq 0$, while rotation-supported systems have properties that are characteristic of late-type galaxies and 
show a broad peak at $\varepsilon \simeq 1$. We define poster child disk galaxies as
systems which, in addition to a characteristic peak at $\varepsilon \simeq 1$, have a
significant cold gas fraction ($f_{\mathrm cold}>0.5$ at $z=2$ and $f_{\mathrm cold}>0.2$ at
$z=0$) with respect to their stellar mass, to distinguish them from transition type systems or ongoing merger events \citep[for details see][]{Teklu15}.
For our simulations it has been shown that, following this classification scheme, galaxies of 
these two populations reproduce accordingly the observed stellar-mass--angular-momentum--relation \citep{Teklu15} and its evolution \citep{Teklu16}, 
the mass-size relation and its evolution \citep{Remus17}, as well as
the fundamental plane distributions \citep{Remus16}.

We then rotate the galaxies such that the minor axis of the gas \footnote{Note that this is different from the computation
for the classification, where the galaxies are rotated into the frame where the angular momentum vector of the stars is
aligned with the $z$-axis.} is aligned with the $z$-axis, so that we can extract the rotation curve without any further
modifications.

From the total of 212 (275) galaxies at $z=2$ ($z=0$) we classify 26 (15)
as poster child disks, which we consider for further analysis. 
In addition, among our 27 poster child spheroidal galaxies
at $z=2$ we find 5 systems with a large cold gas fraction ($f_{\mathrm cold}>0.5$).

Fig.~\ref{fig:maps} shows a 20 kpc region for 4 gas-rich example galaxies at $z=2$, where the
upper row displays the line-of-sight velocity maps of the cold gas component, restricted to regions with $\Sigma_\mathrm{gas} > 50 \frac{M_{\odot}}{pc^2}$, with overlayed cold gas column density contours.
The gridded data was created using SPHMapper (Arth \& Roettgers, in prep.).
The middle row shows the cold gas column density maps with overlayed stellar surface density contours. 
Inclinations and colors were chosen according to the observations presented in \citet{Genzel17}.
Each column represents one galaxy, 
where {\it gal~225}, {\it gal~127}, and {\it gal~62} (from left to right) resemble disk galaxies, 
while {\it gal~183} is a gas-rich spheroidal galaxy. Interestingly, all galaxies, even the spheroidal one, show a similar, regular rotation pattern for the cold gas component.
This is due to the fact that the gas is in a flattened, centrifugally supported disk, even in the systems where
the stars form a spheroid.

The lower panels show mock images of the four galaxies in
the HST broadband F606W (4750A-7000A), which corresponds to rest-frame mid-UV. The images have been generated with the radiative transfer code {\em GRASIL-3D} \citep{Dominguez14}.
This wavelength range traces the regions of very recent star formation, and the spheroidal galaxy shows a very similar mock image as the disks, hiding the real stellar morphology.


\section{Rotation Curves at $z=2$}\label{sec:rotcurve2}

\begin{figure*}
 \begin{centering}
  \centerline{\includegraphics[width=0.49\textwidth,clip=true]{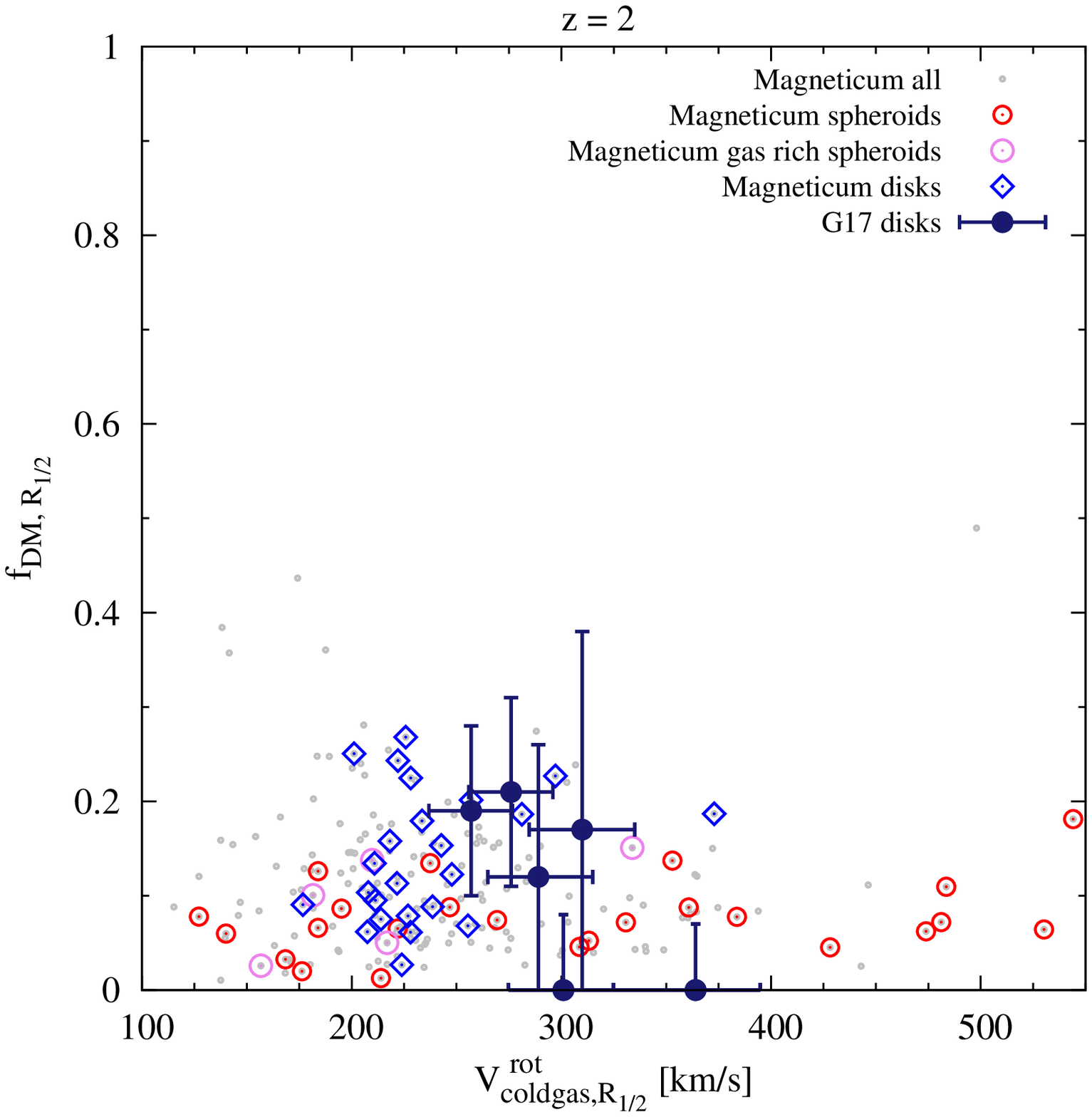} 
              \includegraphics[width=0.49\textwidth,clip=true]{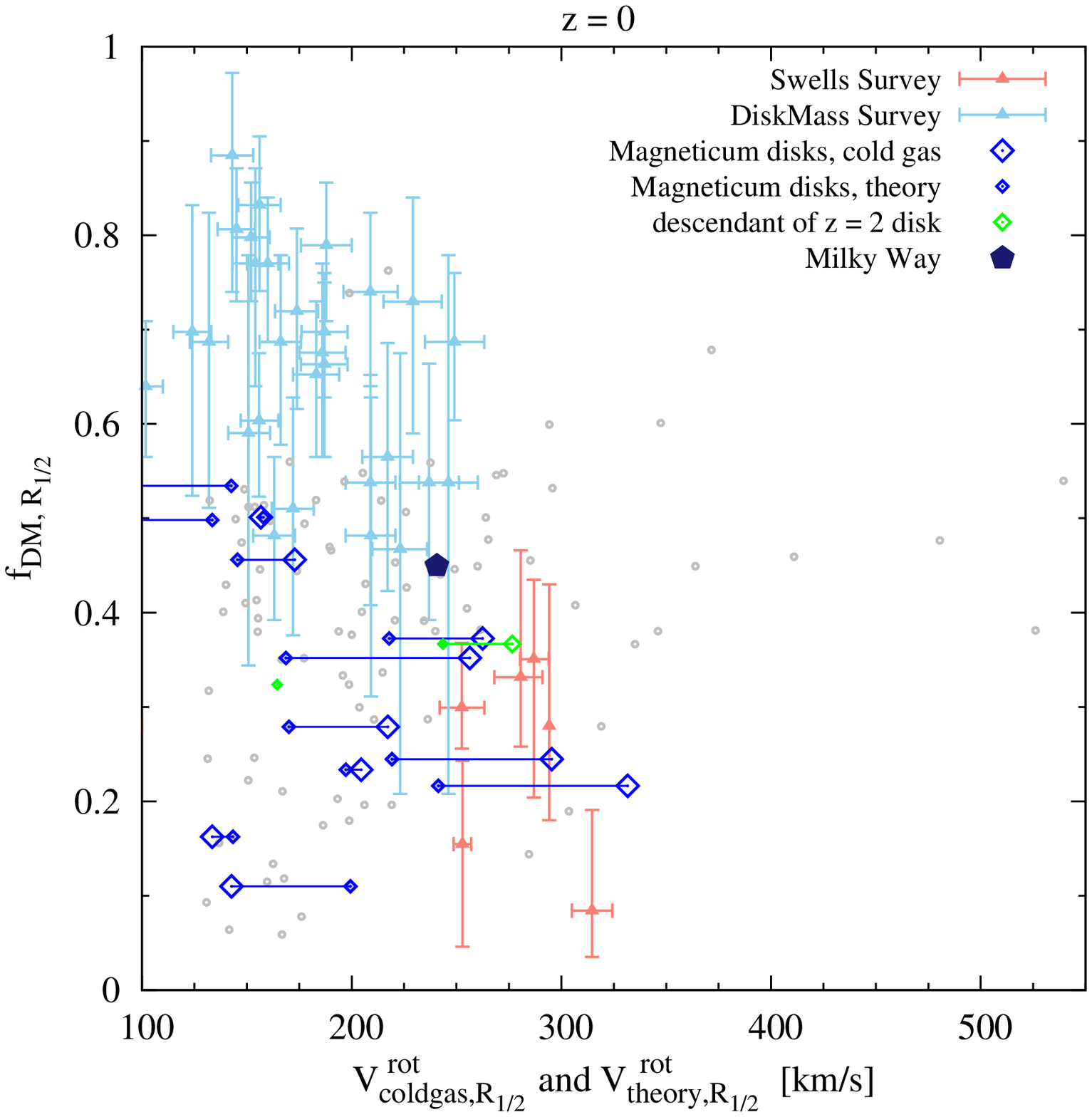}
            }
    \caption{The DM fraction $f_\mathrm{DM}$ within the half-mass radius $R_{1/2}$ versus the rotational velocity $V_\mathrm{coldgas}^\mathrm{rot}$ at $R_{1/2}$ at redshifts $z=2$ (left) and $z=0$ (right).
	At $z=2$ (left panel), the simulated disks (blue diamonds) and gas-rich spheroidals (pink open circles) are shown together with the gas-poor spheroidals (red open circles). 
	The observations from \citet{Genzel17} are included as dark-blue points.
	At $z=0$ (right panel), we only show the simulated disk galaxies, together with observations as presented in \citet{Courteau15} from the Swells Survey \citep{Barnabe12,Dutton13} and the DiskMass Survey \citep{Martinsson13}. 
    The dark-blue filled pentagon shows the Milky Way according to \citet{Bland-Hawthorn16}.
	To indicate uncertainties involved in inferring $V_\mathrm{coldgas,R_{1/2}}^\mathrm{rot}$ we include for the simulated galaxies both the measured rotational gas 
	velocity at $R_{1/2}$ as well as the theoretical value obtained from the total mass within $R_{1/2}$ and connect both points by lines.
        We explicitly highlight the data points for those descendents of our $z=2$ disk galaxies which are
        still disk galaxies at $z=0$ (green diamonds).
    }
  \label{fig:dm_fracs}
 \end{centering}
\end{figure*}

The rotation curves for our galaxy sample are directly obtained from the
averaged velocities (i.e. the circular velocities) of the individual cold gas particles.
In order to ensure that only gas 
within the disk contributes to the rotation curve, only particles
within the $z$-range of $\pm$3kpc are used. While the $z=0$ disk galaxies
show normal rotation curves, 12 out of the 26 poster child disk galaxies at $z=2$
show a significantly declining rotation profile for their gas disk. 
However, we further remove 2 of the 12 examples from our detailed analysis, as they show remnants of recent merger activity.

The left panel of Fig.~\ref{fig:rotcurve2} shows the rotation curves
for these 10 poster child disk galaxies at $z=2$, 
which exhibit a decline in the rotation curve similar to the observed high-$z$ disk galaxies presented in \citet{Genzel17}
(thick solid lines). Following the observations, we scaled the individual
rotation curves by $R_\mathrm{max}$ and $V_\mathrm{max}^\mathrm{rot}$, where $R_\mathrm{max}$
is the radius at which the rotational velocity ($V_\mathrm{coldgas}^\mathrm{rot}$) has its maximum.
We only plot radii larger than two times the gravitational softening of the gas particles, which
corresponds to $\simeq1.33$kpc at $z=2$. As can clearly be seen, the simulated
galaxies show the same behaviour as the observed ones, with some having 
an even
steeper decline in the rotation curves as the observed galaxies.
For comparison, the rotation curves of 7 disk galaxies at $z=0$ are shown as gray lines. 
The difference in profile shapes between high-z and present-day galaxies is clearly visible.

Since at high redshift galaxies are in general more gas-rich, we also plot
the same curves for the 5 gas-rich spheroidal galaxies from our $z=2$ sample
in the right panel of Fig.~\ref{fig:rotcurve2}. As for the disks, the gas shows a clear
rotational pattern (see also example in Fig.~\ref{fig:maps}), and all of our
gas-rich spheroidal galaxies show a declining rotation curve similar to the observed disk galaxies.
The only difference here is that the gas disks in the spheroidals are much smaller than the stellar spheroidal bodies, while the sizes are similar in the disk galaxy cases (see Fig.~\ref{fig:maps}).

The high redshift HST images mainly show young stars, which morphologically closely resemble the gas disks even in the spheroidals (see lower panel of Fig.~\ref{fig:maps}). 
This indicates a potential difficulty in distinguishing disk galaxies from gas-rich spheroidals at $z=2$ observationally. 
However, this uncertainty should be resolved using the next generation of telescopes which will be able to probe the old stellar component in high redshift systems as well.


\section{DM Fractions}\label{sec:DMfrac}
For spheroidal galaxies it is well known that the DM
fraction within the half-mass radius is decreasing at higher
redshift, which is commonly interpreted as indication for late
growth by dry mergers of such systems. While this trend is qualitatively supported by cosmological simulations independent of
the details in the implemented feedback models, the AGN feedback
used in our simulation has been shown to produce DM fractions which
quantitatively agree well with observations \citep[see][]{Remus17}.

The left panel of Fig.~\ref{fig:dm_fracs} shows the DM fractions within 
the stellar half-mass radius $R_{1/2}$ for our full galaxy sample (gray dots) compared
to observations at $z=2$. Generally, our galaxies have a tendency for 
higher average DM fractions with decreasing $V_\mathrm{coldgas}^\mathrm{rot}$, however, nearly all fractions are well below 30\%.
Our disk galaxies (blue diamonds) cover the same range of small DM fractions as the observations presented in \citet{Genzel17}
(dark-blue filled circles with error bars)
\footnote{Note that especially at $z=2$ the unavoidable differences when inferring the half-mass radius in simulations and observations could lead to noticeable differences.}.
Interestingly, the DM fractions of the disk systems are almost as small as those of the spheroidal systems. 
Furthermore, the gas-rich spheroidals cover the same range in DM fractions as the observed and simulated disk galaxies, 
again highlighting the similarities between the gas-rich systems at $z=2$ independent of their morphologies and
demonstrating the difficulty in distinguishing pure rotation-dominated systems from dispersion-dominated systems which host
a significant gas disk.

At $z=0$ the disk galaxies in the simulations show much larger
DM fractions which decrease with rotational velocity
and agree well with the different measurements for disk galaxies (see right panel of Fig.~\ref{fig:dm_fracs}).
To indicate uncertainties involved in inferring $V_\mathrm{coldgas,R_{1/2}}^\mathrm{rot}$ we used both, the measured rotational gas
velocity at $R_{1/2}$ as well as the theoretical values obtained by adopting centrifugal equilibrium and taking the total mass
within $R_{1/2}$.


\begin{figure}
 \begin{centering}
  \centerline{\includegraphics[width=0.49\textwidth,clip=true]{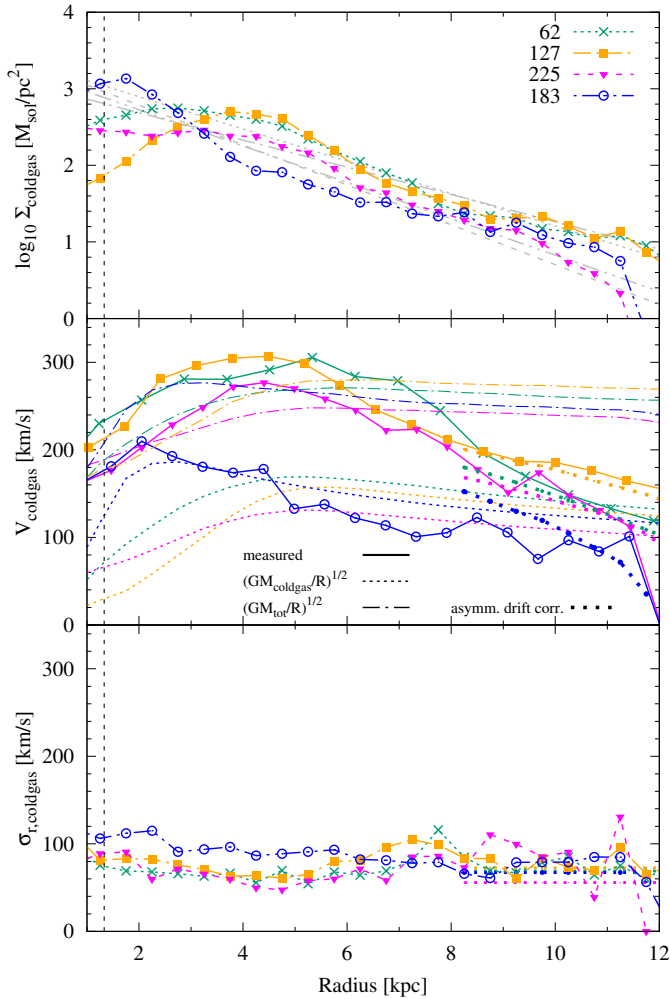}
             }
  \caption{For the three poster child disks ({\it gal~62}, {\it gal~127}, and {\it gal~225}), and the gas-rich spheroidal galaxy ({\it gal~183}): 
    \textit{Upper panel:} Surface density $\Sigma$ of the cold gas. The vertical dashed line indicates four/two times the gravitational
           softening of the stellar/gas particles at this redshift. 
           The gray lines are fits for an exponential surface density profile for $\Sigma(x)=a \cdot exp(-x/b)$ with $b \approx 2$ kpc.
           \textit{Middle panel:} Rotation curves of the cold gas (solid lines) compared to the rotation curves expected from the spherically averaged total mass distribution (dash-dotted lines).
           Dotted lines show the corresponding cold gas contribution.
           The thick dotted lines at large radii show the expected theoretical rotation curves when corrected for the asymmetric drift. 
           \textit{Bottom panel:} Radial velocity dispersion $\sigma_{r}$ of the cold gas. Thin dotted lines indicate the $\sigma$ used for the
           asymmetric drift correction in the middle panel.}
  \label{fig:surfdens_rot}
 \end{centering}
\end{figure}

\section{Surface Density, Dispersion and Theoretical Rotation Curve}\label{sec:surface}
A detailed look at the four examples from Fig.~\ref{fig:maps} shows that the surface density
profiles $\Sigma(r)$ of the cold gas disks in the three poster child disk galaxies and the gas-rich spheroidal galaxy follow the expected exponential decline, as shown in the upper panel of Fig.~\ref{fig:surfdens_rot}. While the
theoretical rotation curves as obtained by the total matter distribution within these halos
are flat, as expected, the real measured rotation of the cold gas disk shows
a significant decline, as can be seen in the middle panel of Fig.~\ref{fig:surfdens_rot}. 
This decline is a result of the kinetic pressure effect which partly compensates the gravitational force as proposed by \citet{Burkert10}.
As expected for a self-gravitating, exponential disk, the maximum of the real rotation curve for the three disk galaxies
in the central part, where the baryons dominate over the dark matter halo, is slightly ($\approx$ 10-20\%) above the maximum
value for a spherically averaged mass distribution \citep{binney}.
Furthermore, at large distances the real rotational velocity is
conspiratorially close to the expected rotational velocity if considering only the cold gas mass. 
For the gas-rich spheroidal galaxy {\it gal 183}, the latter holds even across almost all radii, due to its even lower DM fraction and the small size of the disk compared to the stellar body of the galaxy.
As a result, the gaseous disk of the spheroidal galaxy {\it gal 183} is strongly self-gravitating, more compact
 and shows an even stronger decline. 
None of our systems with a falling rotation curve shows any significant feature or change in the radial component of the
velocity dispersion measured for the cold gas disk which is related to the position at which the rotation curve declines,
as shown in the $\sigma_r$ profiles in the lower panel of Fig.~\ref{fig:surfdens_rot}.

\section{Discussion and Conclusions}\label{sec:conclusion}
Selecting disk galaxies with $M_\mathrm{vir}$ above $5\cdot
10^{11}M_{\odot}$ and $M_*$ above $5\cdot 10^{10} M_{\odot}$ from the
cosmological, hydrodynamical simulation {\it Magneticum Pathfinder} we
investigated the rotation curves of disk galaxies at $z=2$. We find
that almost half of our poster child disk galaxies (10 out of 26) show
significantly declining rotation curves, very similar to the
observations reported in \citet{Genzel17}. Interestingly, the peak of
the rotation curve is a fairly good approximation ($\approx$ 10\%
larger) of the theoretical value, based on the total mass of the
galaxies. 

These disk galaxies do not show any significant dynamical features except that the 
radial dispersion has generally significantly larger values compared to $z=0$ disks,
as expected for dynamically young systems in their assembly phase.
\cite{2012ApJ...754...48F} already presented a model description to explain
this temporal evolution, which also quantitatively agrees well with the results
of the much higher resolution hydrodynamical simulations {\it Eris} \citep{2013ApJ...773...43B},
finding ratios of $~3-4$ between
$v_{\mathrm rot}$ and $\sigma_{\mathrm gas}$ for galaxies at $z=0$ and much smaller
$\sigma_{\mathrm gas}$ for galaxies at $z=0$. Furthermore, this is also
in line with the observational findings of \citet{Simons2017}, who showed that
observed galaxies at $z\sim2$, independently of their stellar mass, typically have
$\sigma_{\mathrm gas}\sim60$ km/s, similar to our galaxies shown in Fig \ref{fig:surfdens_rot}.
Applying a simple correction
$$
  v^2_{\mathrm rot} = v^2_{\mathrm circ} + 2\sigma^2\times(d{\mathrm ln}\Sigma / d{\mathrm ln} R) = v^2_{\mathrm circ} - 2\sigma^2\times(R/R_d)
$$
for the asymmetric drift \citep{Burkert10} based on our measured
dispersion profiles onto the theoretical rotation curve results in
reduced rotation curves, which qualitatively agree well with our
measured ones.  Therefore, we conclude that the declining rotation
curves of the high redshift galaxies are caused by a relatively
thick, turbulent disk, as already discussed in \citet{Genzel17}.  We
also find that these galaxies show similarly low DM fractions as
reported for the observations. The DM halos of these disk galaxies
have a mean concentration parameter $c_\mathrm{vir}\approx8$ (as
expected for these halo masses at $z=2$) and therefore we can
exclude that the low dark matter fractions are caused by especially
low concentrations of the underlying halos.

Tracing these galaxies in the simulations until $z=0$ allows us to
infer the present-day appearances of these galaxies.  We find that, on
average, these galaxies still grow by a factor of $\approx3.5$ both in
virial as well as in stellar mass.  Two of them resemble present-day
disk galaxies with small remaining gas disks, and one ends as a
central galaxy of a small group. Three of them become satellite
galaxies of small groups, while the rest is mostly classified as
transition types.  Therefore, we can exclude that the low DM fractions
at $z=2$ imply that these systems have to be the progenitors of
today's elliptical galaxies with similar stellar mass and low dark
matter fractions.

Interestingly, in our simulations we also find several spheroidal
galaxies at $z=2$ which host a massive cold gas disk with similarly
declining rotation curves as the disk galaxies.  These gas disks are
typically more compact, but  as star formation is dominated by the gas
disks, these spheroidals appear indistinguishable from the disk
galaxies in our mock HST images, highlighting the need for
observational instruments that detect the old stellar component even
at high redshifts.

In general, we conclude that high-redshift disk galaxies with
declining rotation curves and low DM fractions appear naturally within
the $\Lambda$CDM paradigm, reflecting the complex baryonic physics
which plays a role at $z=2$ and can be found commonly in
state-of-the-art, $\Lambda$CDM cosmological hydrodynamical
simulations.


\acknowledgments
We thank Tadziu Hoffmann for useful discussions. 
AFT and KD are supported by the DFG Transregio TR33, AB is supported by the DFG Priority Programme 1573.
AO has been funded by the Deutsche Forschungsgemeinschaft (DFG, German Research Foundation) -- MO 2979/1-1.
This research is supported by the DFG Cluster of Excellence ``Origin and Structure of the Universe.'' 
We are especially grateful for the support by M. Petkova through the Computational Center for Particle and Astrophysics (C2PAP). 
Computations have been performed at the `Leibniz-Rechenzentrum' with CPU time assigned to the Project ``pr86re.''



\begin{thebibliography}{33}
\expandafter\ifx\csname natexlab\endcsname\relax\def\natexlab#1{#1}\fi

\bibitem[{{Abadi} {et~al.}(2003){Abadi}, {Navarro}, {Steinmetz}, \&
  {Eke}}]{Abadi03}
{Abadi}, M.~G., {Navarro}, J.~F., {Steinmetz}, M., \& {Eke}, V.~R. 2003, \apj,
  597, 21

\bibitem[{{Barnab{\`e}} {et~al.}(2012){Barnab{\`e}}, {Dutton}, {Marshall},
  {Auger}, {Brewer}, {Treu}, {Bolton}, {Koo}, \& {Koopmans}}]{Barnabe12}
{Barnab{\`e}}, M., {Dutton}, A.~A., {Marshall}, P.~J., {Auger}, M.~W.,
  {Brewer}, B.~J., {Treu}, T., {Bolton}, A.~S., {Koo}, D.~C., \& {Koopmans},
  L.~V.~E. 2012, \mnras, 423, 1073

\bibitem[{{Beck} {et~al.}(2016){Beck}, {Murante}, {Arth}, {Remus}, {Teklu},
  {Donnert}, {Planelles}, {Beck}, {F{\"o}rster}, {Imgrund}, {Dolag}, \&
  {Borgani}}]{Beck15}
{Beck}, A.~M., {Murante}, G., {Arth}, A., {Remus}, R.-S., {Teklu}, A.~F.,
  {Donnert}, J.~M.~F., {Planelles}, S., {Beck}, M.~C., {F{\"o}rster}, P.,
  {Imgrund}, M., {Dolag}, K., \& {Borgani}, S. 2016, \mnras, 455, 2110

\bibitem[{{Binney} \& {Tremaine}(2008)}]{binney}
{Binney}, J. \& {Tremaine}, S. 2008, {Galactic Dynamics}, 2nd edn. (Princeton
  University Press)

\bibitem[{{Bird} {et~al.}(2013){Bird}, {Kazantzidis}, {Weinberg}, {Guedes},
  {Callegari}, {Mayer}, \& {Madau}}]{2013ApJ...773...43B}
{Bird}, J.~C., {Kazantzidis}, S., {Weinberg}, D.~H., {Guedes}, J., {Callegari},
  S., {Mayer}, L., \& {Madau}, P. 2013, \apj, 773, 43

\bibitem[{{Bland-Hawthorn} \& {Gerhard}(2016)}]{Bland-Hawthorn16}
{Bland-Hawthorn}, J. \& {Gerhard}, O. 2016, \araa, 54, 529

\bibitem[{{Burkert} {et~al.}(2010){Burkert}, {Genzel}, {Bouch{\'e}}, {Cresci},
  {Khochfar}, {Sommer-Larsen}, {Sternberg}, {Naab}, {F{\"o}rster Schreiber},
  {Tacconi}, {Shapiro}, {Hicks}, {Lutz}, {Davies}, {Buschkamp}, \&
  {Genel}}]{Burkert10}
{Burkert}, A., {Genzel}, R., {Bouch{\'e}}, N., {Cresci}, G., {Khochfar}, S.,
  {Sommer-Larsen}, J., {Sternberg}, A., {Naab}, T., {F{\"o}rster Schreiber},
  N., {Tacconi}, L., {Shapiro}, K., {Hicks}, E., {Lutz}, D., {Davies}, R.,
  {Buschkamp}, P., \& {Genel}, S. 2010, \apj, 725, 2324

\bibitem[{{Courteau} \& {Dutton}(2015)}]{Courteau15}
{Courteau}, S. \& {Dutton}, A.~A. 2015, \apjl, 801, L20

\bibitem[{{Dolag} {et~al.}(2016){Dolag}, {Komatsu}, \& {Sunyaev}}]{Dolag16}
{Dolag}, K., {Komatsu}, E., \& {Sunyaev}, R. 2016, \mnras, 463, 1797

\bibitem[{{Dolag} {et~al.}(2017){Dolag}, {Mevius}, \& {Remus}}]{Dolag17}
{Dolag}, K., {Mevius}, E., \& {Remus}, R.-S. 2017, Galaxies, 5, 35

\bibitem[{{Dom{\'{\i}}nguez-Tenreiro}
  {et~al.}(2014){Dom{\'{\i}}nguez-Tenreiro}, {Obreja}, {Granato}, {Schurer},
  {Alpresa}, {Silva}, {Brook}, \& {Serna}}]{Dominguez14}
{Dom{\'{\i}}nguez-Tenreiro}, R., {Obreja}, A., {Granato}, G.~L., {Schurer}, A.,
  {Alpresa}, P., {Silva}, L., {Brook}, C.~B., \& {Serna}, A. 2014, \mnras, 439,
  3868

\bibitem[{{Dutton} {et~al.}(2013){Dutton}, {Treu}, {Brewer}, {Marshall},
  {Auger}, {Barnab{\`e}}, {Koo}, {Bolton}, \& {Koopmans}}]{Dutton13}
{Dutton}, A.~A., {Treu}, T., {Brewer}, B.~J., {Marshall}, P.~J., {Auger},
  M.~W., {Barnab{\`e}}, M., {Koo}, D.~C., {Bolton}, A.~S., \& {Koopmans},
  L.~V.~E. 2013, \mnras, 428, 3183

\bibitem[{{Forbes} {et~al.}(2012){Forbes}, {Krumholz}, \&
  {Burkert}}]{2012ApJ...754...48F}
{Forbes}, J., {Krumholz}, M., \& {Burkert}, A. 2012, \apj, 754, 48

\bibitem[{{Genzel} {et~al.}(2017){Genzel}, {Schreiber}, {{\"U}bler}, {Lang},
  {Naab}, {Bender}, {Tacconi}, {Wisnioski}, {Wuyts}, {Alexander}, {Beifiori},
  {Belli}, {Brammer}, {Burkert}, {Carollo}, {Chan}, {Davies}, {Fossati},
  {Galametz}, {Genel}, {Gerhard}, {Lutz}, {Mendel}, {Momcheva}, {Nelson},
  {Renzini}, {Saglia}, {Sternberg}, {Tacchella}, {Tadaki}, \&
  {Wilman}}]{Genzel17}
{Genzel}, R., {Schreiber}, N.~M.~F., {{\"U}bler}, H., {Lang}, P., {Naab}, T.,
  {Bender}, R., {Tacconi}, L.~J., {Wisnioski}, E., {Wuyts}, S., {Alexander},
  T., {Beifiori}, A., {Belli}, S., {Brammer}, G., {Burkert}, A., {Carollo},
  C.~M., {Chan}, J., {Davies}, R., {Fossati}, M., {Galametz}, A., {Genel}, S.,
  {Gerhard}, O., {Lutz}, D., {Mendel}, J.~T., {Momcheva}, I., {Nelson}, E.~J.,
  {Renzini}, A., {Saglia}, R., {Sternberg}, A., {Tacchella}, S., {Tadaki}, K.,
  \& {Wilman}, D. 2017, \nat, 543, 397

\bibitem[{{Gupta} {et~al.}(2017){Gupta}, {Saro}, {Mohr}, {Dolag}, \&
  {Liu}}]{Gupta17}
{Gupta}, N., {Saro}, A., {Mohr}, J.~J., {Dolag}, K., \& {Liu}, J. 2017, \mnras,
  469, 3069

\bibitem[{{Hirschmann} {et~al.}(2014){Hirschmann}, {Dolag}, {Saro}, {Bachmann},
  {Borgani}, \& {Burkert}}]{Hirschmann14a}
{Hirschmann}, M., {Dolag}, K., {Saro}, A., {Bachmann}, L., {Borgani}, S., \&
  {Burkert}, A. 2014, \mnras, 442, 2304

\bibitem[{{Komatsu} {et~al.}(2011){Komatsu}, {Smith}, {Dunkley}, {Bennett},
  {Gold}, {Hinshaw}, {Jarosik}, {Larson}, {Nolta}, {Page}, {Spergel},
  {Halpern}, {Hill}, {Kogut}, {Limon}, {Meyer}, {Odegard}, {Tucker}, {Weiland},
  {Wollack}, \& {Wright}}]{Komatsu11}
{Komatsu}, E., {Smith}, K.~M., {Dunkley}, J., {Bennett}, C.~L., {Gold}, B.,
  {Hinshaw}, G., {Jarosik}, N., {Larson}, D., {Nolta}, M.~R., {Page}, L.,
  {Spergel}, D.~N., {Halpern}, M., {Hill}, R.~S., {Kogut}, A., {Limon}, M.,
  {Meyer}, S.~S., {Odegard}, N., {Tucker}, G.~S., {Weiland}, J.~L., {Wollack},
  E., \& {Wright}, E.~L. 2011, \apjs, 192, 18

\bibitem[{{Lang} {et~al.}(2017){Lang}, {F{\"o}rster Schreiber}, {Genzel},
  {Wuyts}, {Wisnioski}, {Beifiori}, {Belli}, {Bender}, {Brammer}, {Burkert},
  {Chan}, {Davies}, {Fossati}, {Galametz}, {Kulkarni}, {Lutz}, {Mendel},
  {Momcheva}, {Naab}, {Nelson}, {Saglia}, {Seitz}, {Tacchella}, {Tacconi},
  {Tadaki}, {{\"U}bler}, {van Dokkum}, \& {Wilman}}]{Lang17}
{Lang}, P., {F{\"o}rster Schreiber}, N.~M., {Genzel}, R., {Wuyts}, S.,
  {Wisnioski}, E., {Beifiori}, A., {Belli}, S., {Bender}, R., {Brammer}, G.,
  {Burkert}, A., {Chan}, J., {Davies}, R., {Fossati}, M., {Galametz}, A.,
  {Kulkarni}, S.~K., {Lutz}, D., {Mendel}, J.~T., {Momcheva}, I.~G., {Naab},
  T., {Nelson}, E.~J., {Saglia}, R.~P., {Seitz}, S., {Tacchella}, S.,
  {Tacconi}, L.~J., {Tadaki}, K.-i., {{\"U}bler}, H., {van Dokkum}, P.~G., \&
  {Wilman}, D.~J. 2017, \apj, 840, 92

\bibitem[{{Martinsson} {et~al.}(2013){Martinsson}, {Verheijen}, {Westfall},
  {Bershady}, {Andersen}, \& {Swaters}}]{Martinsson13}
{Martinsson}, T.~P.~K., {Verheijen}, M.~A.~W., {Westfall}, K.~B., {Bershady},
  M.~A., {Andersen}, D.~R., \& {Swaters}, R.~A. 2013, \aap, 557, A131

\bibitem[{{Naab} \& {Ostriker}(2017)}]{Naab17}
{Naab}, T. \& {Ostriker}, J.~P. 2017, \araa, 55, 59

\bibitem[{{Remus} \& {Dolag}(2016)}]{Remus16}
{Remus}, R.-S. \& {Dolag}, K. 2016, in The Interplay between Local and Global
  Processes in Galaxies,

\bibitem[{{Remus} {et~al.}(2017{\natexlab{a}}){Remus}, {Dolag}, \&
  {Hoffmann}}]{Remus17b}
{Remus}, R.-S., {Dolag}, K., \& {Hoffmann}, T. 2017{\natexlab{a}}, Galaxies, 5,
  49

\bibitem[{{Remus} {et~al.}(2017{\natexlab{b}}){Remus}, {Dolag}, {Naab},
  {Burkert}, {Hirschmann}, {Hoffmann}, \& {Johansson}}]{Remus17}
{Remus}, R.-S., {Dolag}, K., {Naab}, T., {Burkert}, A., {Hirschmann}, M.,
  {Hoffmann}, T.~L., \& {Johansson}, P.~H. 2017{\natexlab{b}}, \mnras, 464,
  3742

\bibitem[{{Rubin} {et~al.}(1978){Rubin}, {Thonnard}, \& {Ford}}]{Rubin78}
{Rubin}, V.~C., {Thonnard}, N., \& {Ford}, Jr., W.~K. 1978, \apjl, 225, L107

\bibitem[{{Scannapieco} {et~al.}(2008){Scannapieco}, {Tissera}, {White}, \&
  {Springel}}]{Scan08}
{Scannapieco}, C., {Tissera}, P.~B., {White}, S.~D.~M., \& {Springel}, V. 2008,
  \mnras, 389, 1137

\bibitem[{{Schulze} {et~al.}(2018){Schulze}, {Remus}, {Dolag}, {Burkert},
  {Emsellem}, \& {van de Ven}}]{Schulze2018}
{Schulze}, F., {Remus}, R.-S., {Dolag}, K., {Burkert}, A., {Emsellem}, E., \&
  {van de Ven}, G. 2018, ArXiv e-prints

\bibitem[{{Simons} {et~al.}(2017){Simons}, {Kassin}, {Weiner}, {Faber},
  {Trump}, {Heckman}, {Koo}, {Pacifici}, {Primack}, {Snyder}, \& {de la
  Vega}}]{Simons2017}
{Simons}, R.~C., {Kassin}, S.~A., {Weiner}, B.~J., {Faber}, S.~M., {Trump},
  J.~R., {Heckman}, T.~M., {Koo}, D.~C., {Pacifici}, C., {Primack}, J.~R.,
  {Snyder}, G.~F., \& {de la Vega}, A. 2017, \apj, 843, 46

\bibitem[{{Steinborn} {et~al.}(2016){Steinborn}, {Dolag}, {Comerford},
  {Hirschmann}, {Remus}, \& {Teklu}}]{Steinborn16}
{Steinborn}, L.~K., {Dolag}, K., {Comerford}, J.~M., {Hirschmann}, M., {Remus},
  R.-S., \& {Teklu}, A.~F. 2016, \mnras, 458, 1013

\bibitem[{{Steinborn} {et~al.}(2015){Steinborn}, {Dolag}, {Hirschmann},
  {Prieto}, \& {Remus}}]{Steinborn15}
{Steinborn}, L.~K., {Dolag}, K., {Hirschmann}, M., {Prieto}, M.~A., \& {Remus},
  R.-S. 2015, \mnras, 448, 1504

\bibitem[{{Teklu} {et~al.}(2016){Teklu}, {Remus}, \& {Dolag}}]{Teklu16}
{Teklu}, A.~F., {Remus}, R.-S., \& {Dolag}, K. 2016, in The Interplay between
  Local and Global Processes in Galaxies,

\bibitem[{{Teklu} {et~al.}(2015){Teklu}, {Remus}, {Dolag}, {Beck}, {Burkert},
  {Schmidt}, {Schulze}, \& {Steinborn}}]{Teklu15}
{Teklu}, A.~F., {Remus}, R.-S., {Dolag}, K., {Beck}, A.~M., {Burkert}, A.,
  {Schmidt}, A.~S., {Schulze}, F., \& {Steinborn}, L.~K. 2015, \apj, 812, 29

\bibitem[{{Teklu} {et~al.}(2017){Teklu}, {Remus}, {Dolag}, \&
  {Burkert}}]{Teklu17}
{Teklu}, A.~F., {Remus}, R.-S., {Dolag}, K., \& {Burkert}, A. 2017, \mnras,
  472, 4769

\bibitem[{{Zwicky}(1933)}]{Zwicky33}
{Zwicky}, F. 1933, Helvetica Physica Acta, 6, 110

\end{thebibliography}


\end{document}